\documentclass[prb,showpacs,preprintnumbers,amsmath,amssymb,twocolumn,superscriptaddress,longbibliography]{revtex4-1}

\usepackage[dvipdfmx]{graphicx}
\usepackage{dcolumn}% Align table columns on decimal point
\usepackage{bm}% bold math
\usepackage{color}

\usepackage{amsmath}	% required for `\align' (yatex added)
\usepackage{amsfonts}
\usepackage{braket}
\usepackage{amsthm}

\usepackage{bbm}

\newtheorem{claim}{Claim}

\begin{document}

\title{
Polarization-based indices in quantum many-body systems: validity and extension beyond one dimension
}

\author{Yasuhiro Tada}
\email[]{ytada@hiroshima-u.ac.jp}
\affiliation{
Quantum Matter Program, Graduate School of Advanced Science and Engineering, Hiroshima University,
Higashihiroshima, Hiroshima 739-8530, Japan}
\affiliation{Institute for Solid State Physics, University of Tokyo, Kashiwa 277-8581, Japan}

%\author{Masaki Oshikawa}
%\affiliation{Institute for Solid State Physics, University of Tokyo, Kashiwa 277-8581, Japan}

\begin{abstract}
The expectation value of the twist operator has been widely used as a polarization-based index for gapped and gapless phases in interacting quantum many-body systems. 
%Although numerous 
%studies support this usage in specific settings, the precise conditions determining the validity of such indices in the thermodynamic limit have not always been made explicit.
Although numerous studies support this usage in specific settings and rigorous results have established the validity of the criterion in important settings, the precise assumptions required for it to sharply distinguish gapped and gapless phases under more general conditions have not been fully clarified.
In this work, we clarify the logical status of polarization-based indices by formulating symmetry-based statements under explicitly stated assumptions. 
%We identify the role of ground state degeneracy in gapped systems and show that a unity expectation value is guaranteed under  certain conditions. Independently, we exclude gapless scenarios that could otherwise mimic gapped behavior in the thermodynamic limit.
We identify the role of ground-state degeneracy in the statements for gapped systems and clarify the distinct assumptions required to exclude gapless scenarios that could otherwise mimic gapped behavior in the thermodynamic limit.
%We identify the role of ground-state degeneracy in the statements for gapped systems and clarify which assumptions are required for polarization-based quantities to sharply distinguish gapped and gapless behavior in the thermodynamic limit.
Building on this controlled framework, we construct a meaningful extension beyond one dimension, emphasizing that such an extension is nontrivial and cannot be obtained by a straightforward generalization of the one-dimensional twist operator. Our results delineate the regime in which polarization-based quantities are justified as sharply defined many-body indices.
\end{abstract}

\maketitle

%%%%%%%%%%%%%%%%%%%%%%%%%%%%%%%%%%%%%%%%%%%%%%%%%%%%%%%%%%%%%%%%%%%%%%%%%%
\section{Introduction}
%%%%%%%%%%%%%%%%%%%%%%%%%%%%%%%%%%%%%%%%%%%%%%%%%%%%%%%%%%%%%%%%%%%%%%%%%%

The twist operator has long been employed as a diagnostic for gapped and gapless phases in interacting quantum many-body systems~\cite{Resta1998,RestaSorella1999,Resta2000,Resta2002,Resta2010,
AligiaOrtiz1999,OrtizAligia2000}. 
Since the original proposal, polarization-based quantities have been extensively used in analytical arguments and numerical simulations, particularly in one-dimensional systems, and have become widely accepted tools in the study of correlated quantum systems
\cite{Souza2000,Sgiarovello2001,NakamuraVoit2002,Oshikawa2003,Capello2005,Bendazzoli2010,Stella2011,Varma2015,Yahyavi2017,WatanabeOshikawa2018,Hetenyi2019,Motta2020,Nakamura2002,Kobayashi2018,FuruyaNakamura2019}.

This widespread usage has been supported by a large body of existing work. 
Broadly speaking, earlier studies on the twist operator as a diagnostic may be classified into physical arguments based on single-particle or weakly interacting wavefunctions~\cite{RestaSorella1999,Resta2000}, formal discussions and perturbative calculations in many-body systems~\cite{Resta1998,AligiaOrtiz1999,OrtizAligia2000,Souza2000, WatanabeOshikawa2018}, field-theoretical analyses at low energies~\cite{NakamuraVoit2002,Kobayashi2018,FuruyaNakamura2019}, 
and extensive numerical calculations in specific systems~\cite{RestaSorella1999,Resta2000,Resta2002,Resta2010,AligiaOrtiz1999,OrtizAligia2000,Sgiarovello2001,NakamuraVoit2002,Capello2005,Bendazzoli2010,Stella2011,Varma2015,Yahyavi2017,Hetenyi2019,Motta2020,Nakamura2002,Kobayashi2018}. 
These approaches have provided valuable physical insight and strong evidence for the practical usefulness of the twist operator as a diagnostic. At the same time, they are typically restricted to particular systems, symmetry settings, or controlled limits, and the precise regimes of validity of the criterion, namely, the assumptions required for it to sharply distinguish gapped and gapless phases in the thermodynamic limit, have not always been made explicit.
In this sense, it has remained unclear whether the expectation value of 
the twist operator can serve as a sharp index under fully general conditions.

A major advance was achieved by Tasaki, who provided a general proof for one-dimensional uniquely gapped systems with U(1) symmetry at integer fillings in the thermodynamic limit~\cite{Tasaki2018,Tasaki2022inbook,Tasaki2023}. 
It was rigorously shown that the absolute value of the ground state expectation value of the twist operator converges to unity under well-defined assumptions. This result placed polarization-based diagnostics on a firm mathematical footing and provided a formulation whose underlying structure is extendable to gapped ground states with degeneracy as encountered at non-integer fillings.
A concise reformulation of the essential argument for large but finite systems was later presented~\cite{Tada2024}. %making the structure of the proof more transparent. 
More recently, it was demonstrated that, when an enlarged SO(3) symmetry is imposed, the conclusion can be strengthened further~\cite{Su2025}.  %eliminating the need for certain assumptions on the ground-state structure.
Notably, the position operator entering the twist operator is super-extensive, so that standard exponential clustering arguments for extensive sums of local operators are not directly applicable. 
In this sense, the rigorous control achieved by Tasaki represents a nontrivial step toward a more general theoretical understanding of polarization-based diagnostics.

Even taking these results into account, however, several logical possibilities remained open. 
First, outside the specific settings considered above, it was not excluded that a gapped system could exhibit an expectation value strictly smaller than unity, depending on the detailed structure and degeneracy of the ground state manifold. In particular, the role of ground state degeneracy had not been systematically clarified beyond the simplest cases. 
Second, and independently, the statement that gapped systems yield a unity expectation value does not by itself exclude the logical possibility that certain gapless systems might also have a non-zero or unity expectation value of the twist operator in the thermodynamic limit. 
Clarifying which assumptions are required to rule out these possibilities is essential for establishing polarization-based quantities as sharply defined many-body indices, rather than phenomenological diagnostics whose validity is inferred only a posteriori.

Additional subtleties arise when one attempts to extend polarization-based diagnostics beyond one dimension. It is now well understood that a straightforward generalization of the one-dimensional twist or polarization operator is delicate in higher-dimensional systems. 
A closely related lesson is provided by higher-dimensional generalizations of the Lieb-Schultz-Mattis theorem~\cite{LSM1961}. In contrast to one dimension, such extensions require intrinsically higher-dimensional ingredients, such as flux insertion and locality estimates, and cannot be obtained by a naive transposition of the original one-dimensional argument~\cite{Oshikawa2000_1,Hastings2004,NachtergaeleSims2007,BachmannBolsDeRoeckFraas2020}.
For the polarization, in particular, it was shown that naive generalization of the polarization operator can fail: for example, when higher-dimensional systems are constructed by stacking or bundling one-dimensional chains, the thermodynamic limit behavior of the many-body polarization amplitude can become nontrivial and may yield a vanishingly small value %$|\braket{U}|<1$
even in settings where a one-dimensional intuition might suggest convergence to unity~\cite{WatanabeOshikawa2018,Nakagawa2017,Misawa2023}. 
This illustrates that the extension of polarization-based diagnostics to higher dimensions cannot be justified by a purely formal transposition of one-dimensional expressions and must be treated with care.
Related insights have also been obtained from studies of multipole moments. 
Formal extensions from dipole polarization to higher multipole quantities such as quadrupole moments were shown not to yield a simple binary (0/1) index in general higher dimensional systems~\cite{Kang2019,Wheeler2019,Ono2019,Tada2023}, 
underscoring again that polarization-based constructions become increasingly subtle beyond one dimension. 
Even the formulation of electric polarization which is the lowest order multipole in higher-dimensional systems is non-trivial and has remained an active subject of recent research~\cite{Song2021,Zhang2023}.
These developments indicate that the problem of defining a meaningful polarization-based index in higher dimensions is itself nontrivial and cannot be addressed by purely formal extensions of one-dimensional expressions. Rather, one must carefully identify the assumptions under which such quantities are well defined and determine how symmetry and low energy spectral properties constrain their behavior in the thermodynamic limit.

In this work, we address these issues by formulating symmetry-based statements that directly control the thermodynamic limit of the twist operator under explicitly stated conditions. We clarify the assumptions required for polarization-based quantities to exhibit a sharp 0/1 dichotomy, identify the role of ground state degeneracy in gapped systems, and independently exclude gapless scenarios that would otherwise mimic gapped behavior. Building on this controlled framework, we then construct a meaningful extension of polarization-based indices beyond one dimension that is consistent with the known subtleties of higher-dimensional polarization. Although our results are established at the level of general statements and proofs, we include numerical calculations in Appendix as illustrative demonstrations. These calculations are not required for the logical completeness of the arguments, but serve to connect the general framework to familiar settings and to confirm consistency with existing numerical observations. Our results place polarization-based diagnostics on a firmer logical footing and delineate the regime in which their use as many-body indices is justified.

%%%%%%%%%%%%%%%%%%%%%%%%%%%%%%%%%%%%%%%%%%%%%%%%%%%%%%%%%%%%
\section{Main results and proofs}
%%%%%%%%%%%%%%%%%%%%%%%%%%%%%%%%%%%%%%%%%%%%%%%%%%%%%%%%%%%%
%%%%%%%%%%%%%%%%%%%%%%%%%%%%%%%%%%%%%%%%%%%%%%%%%%%%%%%%%%%%
\subsection{Claims and Remarks}
%%%%%%%%%%%%%%%%%%%%%%%%%%%%%%%%%%%%%%%%%%%%%%%%%%%%%%%%%%%%

As a prototypical model,
we consider one dimensional U(1) symmetric spinless particles on a chain with a single-site unit cell described by the Hamiltonian
%%%%%%%%%%%%%%%%%%%%%%%%%%%
\begin{align}
H=\sum_{jk}t_{jk}c^{\dagger}_jc_k + \sum_{jk}V_{jk} n_j n_k.
\label{eq:H}
\end{align}
%%%%%%%%%%%%%%%%%%%%%%%%%
$c_{j}$ is the annihilation operator of a fermion or boson 
at a site $j$ and $n_j=c^{\dagger}_jc_j$ is the charge density
operator.  %with the filling $\nu$ per site.
For a one-dimensional system with length $L$ and the number of the particles $N$, 
the filling per site $\nu=N/L=p/q$ is a conserved quantity
where $p,q$ are coprime.
We first consider one dimensional systems and generalization to higher dimensions will be discussed later.
$t_{jk}$ is the hopping and $V_{jk}$ is the interaction strength.
We assume that ranges of $t_{jk}$ and $V_{jk}$ are finite, and for simplicity, we consider $t_{jk}=t$ is non-zero
only for the nearest neighbor sites.
It is straightforward to extend the following discussions to a system with further neighbor hoppings and also with a unit cell containing
multiple sites.
A system with a unit cell containing multiple sites can be discussed similarly to a model with a single site unit cell.
The site position is denoted by $j=J+a$, where $J=0,1,\cdots,L-1$ is the position of the unit cell and $a$ is the position within the unit cell.
Filling per unit cell is given by $\nu=N/L=p/q$.
Note that one can also consider a spin-$S$ system described by the Hamiltonian
%%%%%%%%%%%%%%%%%%%%%%%%%%%
\begin{align}
H=\sum_{jk} \frac{J_{jk}}{2}(S^+_jS^-_k+S^-_jS^+_k)+J^z_{jk}S^z_jS^z_k,
\end{align}
%%%%%%%%%%%%%%%%%%%%%%%%%
where the spin $S_j^z$ operator plays the same roles as those of $n_j$ and magnetization $m=\sum S_j^z/L$ corresponds to the filling $\nu$.
%The spin model can be mapped to a hard-core boson model Eq.~\eqref{eq:H}, and thus our results are applicable also to spin systems.
Our results apply also to spin systems with an appropriate identification of $S^z_j$ and $n_j$.

We impose the periodic boundary condition and suppose that the system has the translation symmetry,
$[H,T]=0$, where $T$ is the translation operator $Tc_{j}T^{-1}=c_{j+1}$.
%Since $T\hat{c}_kT^{-1}=e^{-i2\pi k/L}\hat{c}_k$ for the Fourier transformed operator 
%$\hat{c}_k=(1/\sqrt{L})\sum_jc_je^{-i2\pi kj/L}$ $(k=0,1,\cdots,L-1)$,
$T$ is explicitly expressed as 
%%%%%%%%%%%%%%%%%%%%%%%%%%%
\begin{align}
T=\exp \left( i\frac{2\pi}{L}\sum_{k=0}^{L-1} k\hat{n}_k\right),
\label{eq:T}
\end{align}
%%%%%%%%%%%%%%%%%%%%%%%%%
where $\hat{c}_k$ is the Fourier transformation of $c_j$ and $\hat{n}_k=\hat{c}^{\dagger}_k\hat{c}_k$.
We also introduce the dual operator
%%%%%%%%%%%%%%%%%%%%%%%%%%%
\begin{align}
U=\exp \left( i\frac{2\pi}{L}\sum_{j=0}^{L-1} jn_j\right),
\label{eq:U}
\end{align}
%%%%%%%%%%%%%%%%%%%%%%%%%
which transforms the operators as $Uc_j^{\dagger}U^{-1}=e^{i2\pi j/L}c_j^{\dagger}$.
This is the polarization operator~\cite{Resta1998,RestaSorella1999} and is also called the Lieb-Schultz-Mattis twist operator~\cite{LSM1961,AffleckLieb1986,OYA1997,YOA1997,Tasaki2022inbook}.
These two unitary operators satisfy the commutation relation,
%%%%%%%%%%%%%%%%%%%%%%%%%%%
\begin{align}
UTU^{-1}=e^{i2\pi\nu}T
\label{eq:UT}
\end{align}
%%%%%%%%%%%%%%%%%%%%%%%%%
which is similar to the Weyl representation of the canonical commutation relation in quantum mechanics.
%For a set of $q$ states with the momentum eigenvalues $\kappa^k=e^{i2\pi\nu k}$ $(k=0,1,\cdots,q-1)$,
%%$\{1, e^{i2\pi\nu},\cdots,e^{i2\pi\nu(q-1)}\}$,
%these two operators may be represented as $q\times q$ matrices,
%%%%%%%%%%%%%%%%%%%%%%%%%%%
%\begin{align}
%T=
%\begin{bmatrix}
%  1 & 0 & \dots & 0 \\
%   0 & \kappa & \dots & 0 \\
%   \vdots & \vdots & \ddots & \vdots \\
%   0 & \dots & \dots & \kappa^{q-1}
%\end{bmatrix},
%U=
%\begin{bmatrix}
%   0 & 1 & 0 & \dots & 0 \\
%   0 & 0 & 1 & \dots & 0 \\
%   \vdots & \vdots & \ddots & \ddots & \vdots \\
%   0 & 0 & \dots & 0 & 1 \\
%   1 & 0 & \dots & \dots & 0
%\end{bmatrix}.
%\label{eq:TUmat}
%\end{align}
%%%%%%%%%%%%%%%%%%%%%%%%%
%It turns out that this representation is essentially valid also
%in the gapped ground state sector of the Hamiltonian $H$ with small corrections $O(1/L)$ which vanish
%in the thermodynamic limit.

We consider gapped systems and gapless states respectively.
For gapped systems,
we suppose that the system has quasi-degenerate gapped ground states $\ket{\Psi_{0}},
\ket{\Psi_{1}},\cdots,\ket{\Psi_{D-1}}$, where $D$ is the ground state degeneracy which is assumed to be
independent of the system size $L$.
It is also assumed that energy difference between these quasi-degenerate ground states in a finite size system
is at most $O(1/L)$, namely, $E_{0}\leq E_{1}\leq \cdots\leq E_{D-1}$ with $E_{D-1}-E_{0}=O(1/L)$.
These low energy states are separated from the excited states by a non-zero energy gap $\Delta_0$.
In gapless systems, there is no clear distinction between the ground state sector and the excited states in a finite size system,
since there is no excitation gap.
In the present study, we suppose that the gapless ground state for finite $L$ is non-degenerate, $D=1$.

We first discuss gapped states.
It is stressed that our statements hold also in higher dimensions with appropriate modifications
as will be discussed later.
The following statement is a well-known fundamental property~\cite{LSM1961,AffleckLieb1986}
which holds for both gapped and gapless states.
%%%%%%%%%%%%%%%%%%%%%%%%%%%
\begin{claim}{}
The difference between the variational energy of $\ket{\Phi_{k}}=U\ket{\Psi_{k}}$ and the ground state energy
$E_{k}$ is vanishing in the thermodynamic limit, 
$\Delta E_{k}=\braket{\Phi_{k}|H|\Phi_{k}}-\braket{\Psi_{k}|H|\Psi_{k}}=O(1/L)$
for $k=0,\cdots,D-1$.
%Further, absolute values of any matrix elements $\Delta E_{kk'}=\braket{\Phi_{k}|H|\Phi_{k'}}
%-\braket{\Psi_{0k}|H|\Psi_{0k'}}$ are $O(1/L)$.
\label{claim:dE}
\end{claim}
%%%%%%%%%%%%%%%%%%%%%%%%%
Based on Claim~\ref{claim:dE}, we can show that $U$ works essentially as a unitary rotation within
the quasi-degenerate ground states.
This means that $U$ can be regarded almost as a symmetry within the ground state sector of a gapped system.
%This means that Eq.~\eqref{eq:TUmat} is valid in an approximate sense.
%%%%%%%%%%%%%%%%%%%%%%%%%%%
\begin{claim}{}
When the system is gapped with the quasi-degeneracy $D$, 
the overlap between the variational state $\ket{\Phi_{k}}=U\ket{\Psi_{k}}$ and the ground state $\ket{\Psi_{k}}$,
$Z_{kk'}=\braket{\Psi_{k}|U|\Psi_{k'}}$ for $k,k'=0,\cdots,D-1$, is  
almost a $D\times D$ unitary matrix with $|\det (Z)|=1+O(1/L)$.
\label{claim:detZ}
\end{claim}
%%%%%%%%%%%%%%%%%%%%%%%%%
Claim~\ref{claim:detZ} applies to insulators with an integer $\nu$ ($D=1$) and also a fractional $\nu$ ($D>1$).
The former is typically (correlated) band insulators, 
while the latter includes charge-density-wave orders and intrinsic topological orders.
Note that translation symmetry is not necessary for the statement and Claim~\ref{claim:detZ} holds even for
a system with randomness.
However, an excitation gap is required in the condition, and hence Claim~\ref{claim:detZ} is 
not applicable to Anderson insulators with gapless excitations.
Claim~\ref{claim:detZ} may be useful in exact diagonalization calculations where several lowest energy states
can be obtained with reasonable computational costs.
In a generic system, the first excited state $\ket{\Psi_D}$ would not be an almost eigenstate of $U$.
Then,
if we consider a larger $(D+1)\times(D+1)$ matrix $\tilde{Z}_{kk'}=z_{kk'}$ for 
$k,k'=0,1,\cdots,D-1,D$, the added elements are $z_{k,D}=O(1/L)$ and $\tilde{Z}$ will no longer be almost unitary,
and consequently $|\det(\tilde{Z})|=O(1/L)$.
%since the rank of $\tilde{Z}$ is $D$ but not $D+1$ in the limit $L\to\infty$.
This enables one to evaluate quasi-degeneracy in numerical calculations, as demonstrated in Appendix~\ref{app:1d}.
However, in some other numerical methods such as the density matrix renormalization group method
~\cite{White1992,ITensor2022,TeNPy2018} and 
the many-variable variational Monte Carlo method~\cite{Misawa2019}, additional calculations are necessary to obtain 
quasi-degenerate ground states.
It is desirable to improve Claim~\ref{claim:detZ} so that it involves only a single ground state. 
Under additional conditions, we have the following statement.
%%%%%%%%%%%%%%%%%%%%%%%%%%%
\begin{claim}{}
Suppose that the system has the translation symmetry and the filling is $\nu=p/q$.
When the system is gapped with the quasi-degeneracy $D=q$, 
$|\braket{\Psi_{k}|U^q|\Psi_{k'}}|=\delta_{kk'}+O(1/L)$ in the ground state sector.
Especially, $|\braket{\Psi_{0}|U^q|\Psi_{0}}|=1+O(1/L)$.
\label{claim:Uq}
\end{claim}
%%%%%%%%%%%%%%%%%%%%%%%%%
Note that,
due to the commutation relation Eq.~\eqref{eq:UT},
two states $\ket{\Psi_{k}}$ and $U^l\ket{\Psi_{k}}$ ($l=1,2,\cdots,q-1$) are orthogonal under the translation
symmetry.
On the other hand, orthogonality of $\ket{\Psi_{k}}$ and $U^q\ket{\Psi_{k}}$ is non-trivial,
since they have the same $T$-eigenvalue due to $[T,U^q]=0$.
In Claim~\ref{claim:Uq}, we have assumed that the degeneracy is $D=q$ which is the minimal degeneracy
imposed by the translational symmetry, and consequently the $T$-eigenstates are unique in each momentum sector.
Although $D=q$ will hold if there are neither additional symmetry protected degeneracy nor accidental degeneracy,
unfortunately, it cannot be known a priori in practical calculations.
%If the degeneracy is $D>q$, $U^q\ket{\Psi_{0}}$ might not be an almost eigenstate of $U^q$
%and $|\braket{\Psi_{0}|U^q|\Psi_{0}}|<1$ would be possible.
Such an additional condition is not necessary when the system has SO(3) symmetry which is larger than U(1), 
where each momentum sector has a unique (spin singlet) ground state~\cite{Su2025}.

The above statements claim that non-zero expectation values of $U^q$ signal gapped %(charge gapped) 
nature of the system.
On the other hand,
for gapless systems, the absence of a charge gap (defined in Eq.~\eqref{eq:gap_c}) leads to the following statement.
%%%%%%%%%%%%%%%%%%%%%%%%%%%
\begin{claim}{}
Suppose that the system has the translation symmetry and the filling is $\nu=p/q$. 
%When the system is charge gapless and the ground state is non-degenerate, 
When the ground state is charge gapless and non-degenerate in the finite size system, 
%the ground state expectation value is  
$|\braket{\Psi_{0}|U^q|\Psi_{0}}|\to 0$ in the thermodynamic limit.
\label{claim:Uq_gapless}
\end{claim}
%%%%%%%%%%%%%%%%%%%%%%%%%
This may be an expected property for a gapless system because the polarization given by the phase of $\braket{U^q}$ should not be well-defined, 
but it is highly non-trivial for general interacting systems~\cite{RestaSorella1999,AligiaOrtiz1999,Kobayashi2018,FuruyaNakamura2019}.
Our statement clarifies the explicit condition (vanishing charge gap) for $|\braket{U^q}|\to0$ and 
we will later discuss systems where the condition is not satisfied.  
Under the condition of the vanishing charge gap, Claim~\ref{claim:Uq_gapless} and its higher dimensional generalization 
hold for general gapless systems including Tomonaga-Luttinger liquids, non-Fermi liquids, and spin liquids.
We note that, although the twist operator itself acts within a fixed U(1) sector, the argument crucially exploits the vanishing of the inter-sector (charge) gap as will be discussed in the next section.
Besides,
the argument naturally suggests an algebraic decay of $|\braket{U^q}|$ with the system size $L$ for gapless states.

The above statements mean that $U^{(q)}$ works as a quantized order parameter 
which clearly distinguish between gapped
and gapless states in general interacting systems.
Alternatively, one can regard it as a many-body index %${\rm ind}=\lim_{L\to\infty} |\braket{U^{(q)}}|\in \{0,1\}$, 
for gapped and gapless states,
%%%%%%%%%%%%%%%%%%%%%%%%%%%
\begin{align}
{\rm ind}=\lim_{L\to\infty} |\braket{\Psi_0|U^{q}|\Psi_0}|\in \{0,1\}.
\label{eq:index}
\end{align}
%%%%%%%%%%%%%%%%%%%%%%%%%
This means that existence or absence of an excitation gap can be characterized by an index under the U(1) and translation symmetries
(and under the assumption on the ground state degeneracy in Claims~\ref{claim:Uq} and \ref{claim:Uq_gapless}),
although it is usually considered that symmetries are irrelevant in the classification of gappedness and gaplessness~\cite{Wen2004,Zeng2019}.
%We stress that this hold true even in higher dimensions as will be discussed later.
It is known that ground states of generic local Hamiltonians including non translationally symmetric systems are gapless~\cite{Movassagh2017}, 
which also implies that distinction between gapped and gapless states are meaningful only when the system has sufficient symmetries.
As noted above,
the quantity $|\braket{U^q}|$ itself was originally introduced in the context of polarization-based criteria~\cite{RestaSorella1999,AligiaOrtiz1999}. 
However, prior to the present work, it had not been established as a general and rigorous index which
universally converges to either 0 or 1 and  sharply distinguishes gapped and gapless phases in the thermodynamic limit.

%We present esveral remarks below.
We make several remarks in order to clarify the scope and implications of the above statements.
First, our statements are extendable to spinful particles. 
If there is $\mathrm{U}(1)_{\uparrow}\times\mathrm{U}(1)_{\downarrow}$ symmetry,
they hold for each spin sector.
Alternatively, one can introduce two polarization operators $U_c=e^{i2\pi/L\sum x_jn_j}$ and 
$U_s=e^{i2\pi/L\sum x_js_j^z}$ with $n_j=n_{j\uparrow}+n_{j\downarrow}$ and $s_j^z=(n_{j\uparrow}-n_{j\downarrow})/2$
corresponding to the $\mathrm{U}(1)_{N}\times\mathrm{U}(1)_{S^z}$ symmetry.
They satisfy $TU_cT^{-1}=e^{-i2\pi \nu}U_c$ with $\nu=\nu_{\uparrow}+\nu_{\downarrow}$ and $TU_s^2T^{-1}=e^{-i4\pi m^z}U_s^2$
in a given $m^z=1/L\sum_js_j^z$ sector.
Claims 2 and 3 hold if the system is gapped, and Claim 4 holds for each of the charge and spin gaps.

Second, Claim 3 implies that $\log |\braket{U^q}|$ vanishes at least as $O(1/L)$.
If one formally interprets $\log \braket{U^q}$
in terms of a cumulant expansion of the many-body position operator, this behavior is consistent with an extensive scaling of the second cumulant. %as assumed in the localization-length picture of Resta and Sorella.
We emphasize, however, that Claim 3 itself does not rely on such an expansion.
While Claim 3 only implies $|\braket{U^q}|=1+O(1/L)$, %in many models the finite-size data are well captured by an empirical form 
one can expect an empirical form $|\braket{U^q}|=e^{-\lambda/L}$ with a localization length $\lambda$~\cite{RestaSorella1999,Resta2000,Resta2002,Resta2010,Souza2000}.
Such an exponential scaling can be motivated when $\log \braket{U^q}$ is dominated by the leading cumulant(s).

Third, it is worth emphasizing that
Claims 2 and 3 rely only on the existence of a finite excitation gap within a fixed particle-number sector and on symmetry considerations, and therefore remain valid even in the presence of long-range density–density interactions. 
In contrast, Claim 4 involves comparisons between different particle-number sectors and is restricted to short-range interacting systems.
More specifically,
throughout this work, we consider a fixed Hamiltonian that is independent of the particle-number sector, and the ground state energy $E_0^N$
is defined within the $N$-particle sector of the same Hamiltonian.
Under this convention, the condition of a vanishing charge gap is an additional physical assumption.
For systems with short-range interactions, such a situation naturally arises in compressible phases, where $E_0^N$
is a smooth function of the charge density and the charge gap (Eq.~\eqref{eq:gap_c}) will vanish 
in the thermodynamic limit.
In contrast, in the presence of long-range Coulomb interactions, the interpretation of the charge gap
may depend sensitively on how charge neutrality is implemented across different particle-number sectors.
We have therefore restrict our discussion to short-range interacting systems, for which the charge gap
directly reflects the energy cost of local charge excitations.

Finally,
it should be stressed that the condition in Claim~\ref{claim:Uq_gapless} is the vanishing charge gap.
We can explicitly rule out gapless scenarios that could otherwise mimic gapped behavior under this condition, 
%but there may be a gapless state which does not satisfy the condition and exhibits $\braket{U^q}\neq 0$. 
but there may be a gapless state which does not satisfy the condition and exhibits a gapped like behavior of $\braket{U^q}$.
Generally, there are two kinds of gaps in a U(1) symmetric system, namely an inter-U(1)-sector gap for multiple subspaces with 
different U(1) charges and an intra-U(1)-sector gap within a single subspace with a fixed U(1) charge.
%In the present study, a gapped system has both of an inter-sector gap or an intra-sector gap, while a gapless system has a inter-sector 
One may expect that Claim~\ref{claim:Uq_gapless} can be extended to 
a ground state with a non-zero inter-sector gap and intra-sector gapless excitations.
However, the expectation value can be $|\braket{U^q}|\neq 0$ in some systems satisfying this condition (i.e. inter-sector gapped and intra-sector gapless).
Therefore, such an extension of Claim~\ref{claim:Uq_gapless} is impossible.

It is important to specify which gap is vanishing to establish Eq.~\eqref{eq:index} as a sharply quantized index.
While trivial constructions based on decoupled subsystems can obviously realize gapless energy spectra with gapped like behavior of $\braket{U^q}$, the physically more relevant question is whether such behavior can occur in a system which is not trivially decoupled.
An example is the strong interaction limit of the one-dimensional Hubbard model at half-filling, where the system is effectively described by
the Heisenberg Hamiltonian $H_{\rm eff}$.
The inter-sector gap (charge gap for the particle number U(1)) is non-zero and the intra-sector gap is vanishing due to the spin degrees of freedom.
The polarization operator for the particle number $U_c$ and the effective Hamiltonian commutes, $[U_c,H_{\rm eff}]=0$,
and therefore $|\braket{\Psi_0|U_c|\Psi_0}|=1+$(contributions from high energy states beyond the effective Hamiltonian) and it converges
to unity in the infinitely strong interaction limit, $|\braket{\Psi_0|U_c|\Psi_0}|\to1$.
On the other hand, the spin gap is zero and thus $|\braket{\Psi_0|U_s^2|\Psi_0}|\to0$ for the spin degrees of freedom according to Claim~\ref{claim:Uq_gapless}.
Another example is the 1/3 magnetization plateau in
the $S=1/2 $ three-leg spin tube in the limit $J\gg J'$, where $J$ in the rung interaction and $J'$ is the leg interaction~\cite{Okunishi2012}.
In this system, there is only one U(1) symmetry in contrast to the Hubbard model, and the spin gap is non-zero at the magnetization plateau.
The low energy degrees of freedom are the chirality $\ket{+}, \ket{-}$ in the triangles
and the system is described by an effective spin-1/2 model, leading to the gapless chirality liquid within the fixed U(1) sector~\cite{Okunishi2012}. 
The corresponding effective Hamiltonian $H_{\rm eff}$ commutes with $U=e^{2\pi/L\sum jT^z_j}$, where $T^z_j=\sum_{k=1}^3S^z_{k,j}$ is
the total magnetization for the triangle at the position $j$ in the leg direction.
Since the ground state is almost an eigenstate of $U$, we have $|\braket{\Psi_0|U|\Psi_0}|\simeq1$ for the weak leg interaction $J\gg J'$.
In these examples, the essential point is that
the operator $U^q$ does not create low energy excitations and thus the state $U^q\ket{\Psi_0}$ stays almost at the
ground state $\ket{\Psi_0}$, which results in $|\braket{\Psi_0|U^q|\Psi_0}|\to1$ even though the system is gapless in other degrees of freedom.
Claim~\ref{claim:Uq_gapless} sharpens the statement of the vanishing expectation value $\braket{U^q}$ by explicitly identifying the assumptions under which gapped like behavior is ruled out. %thereby clarifying the logical scope of earlier arguments.

%%%%%%%%%%%%%%%%%%%%%%%%%%%%%%%%%%%%%%%%%%%%%%%%%%%%%%%%%%%%%%%%%%%%%
\subsection{Proofs of Claims}
%%%%%%%%%%%%%%%%%%%%%%%%%%%%%%%%%%%%%%%%%%%%%%%%%%%%%%%%%%%%%%%%%%%%%
In this section, we provide proofs for our Claims for one-dimensional systems. 
Generalization to higher dimensions
will be discussed later in Sec.~\ref{sec:highD}.

%%%%%%%%%%%%%%%%%%%%%%%%%%%%%%%%%%%%%%%%%%%%%
{\it Proof of Claim~\ref{claim:dE}.}
The proof is well-known, but we provide a simple one to be self-contained.
Let us consider the variational states $\ket{\Phi_k}=U\ket{\Psi_{k}}$ $(k=0,1,\cdots,D-1)$
and the energy difference $\Delta E_k=\braket{\Phi_k|H|\Phi_k}-\braket{\Psi_{k}|H|\Psi_{k}}$~\cite{LSM1961,AffleckLieb1986}.
Since $U^{-1}c^{\dagger}_jc_{j+1}U=e^{i2\pi/L}c^{\dagger}_jc_{j+1}=(1+i2\pi/L+\cdots)c^{\dagger}_jc_{j+1}$, we have 
$U^{-1}HU-H=J+O(1/L)$ where
%%%%%%%%%%%%%%%%%%%%%%%%%%%
\begin{align}
J=\frac{2\pi}{L}\sum_j it \left(c^{\dagger}_{j}c_{j+1}-c^{\dagger}_{j+1}c_j\right)
\label{eq:J}
\end{align}
%%%%%%%%%%%%%%%%%%%%%%%%%
is the current density operator (multiplied by $2\pi$).
The variational energy is 
$\Delta E_k=\braket{\Psi_{k}|J|\Psi_{k}}+O(1/L)$.
Now we introduce another ground state $\ket{\Phi_k'}=U^{-1}\ket{\Psi_{k}}$ and evaluate its variational energy
to obtain $\Delta E_k'=\braket{\Psi_{k}|-J|\Psi_{k}}+O(1/L)$.
Since any variational energy must not be smaller than the (lowest) ground state energy $E_{0}$, 
we have $\braket{\Psi_{k}|\pm J|\Psi_{k}}=O(1/L)$, which is known as Bloch-Bohm's theorem
for the absence of a persistent current~\cite{Bohm1949,TadaKoma2016}.
As a result, $\Delta E_k=O(1/L)$.

%%%%%%%%%%%%%%%%%%%%%%%%%%%%%%%%%%%%%%%%%%%%%
{\it Proof of Claim~\ref{claim:detZ}.}
We expand the variational states in terms of the energy eigenstates $\ket{\Psi_{0}},\cdots,\ket{\Psi_{D-1}},
\ket{\Psi_D},\cdots$,
%%%%%%%%%%%%%%%%%%%%%%%%%%%
\begin{align}
\ket{\Phi_k}=U\ket{\Psi_{k}}=\sum_{k=0}^{D-1}z_{kk'}\ket{\Psi_{k'}}+\sum_{n\geq D}z_{kn}\ket{\Psi_n}.
\end{align}
%%%%%%%%%%%%%%%%%%%%%%%%%
Since $U$ is unitary, the variational states are orthonormal, 
%%%%%%%%%%%%%%%%%%%%%%%%%%%
\begin{align}
\braket{\Phi_k|\Phi_{k'}}=\delta_{kk'}=\sum_{l=0}^{D-1}z_{kl}^*z_{k'l}+\sum_{n\geq D}z_{kn}^*z_{k'n}.
\end{align}
%%%%%%%%%%%%%%%%%%%%%%%%%
Now the energy difference $\Delta E_{k}$ for $k=0,1,\cdots, D-1$ is expressed as
%%%%%%%%%%%%%%%%%%%%%%%%%%%
\begin{align}
\Delta E_{k}
%&=\sum_{l=0}^{D-1}z_{kl}^*z_{k'l}E_{0l}+\sum_{n\geq1}z_{kn}^*z_{k'n}E_n -E_{0k}\delta_{kk'} \nonumber\\
&=\sum_{l=0}^{D-1}|z_{kl}|^2(E_{l}-E_{k})+\sum_{n\geq D}|z_{kn}|^2(E_n-E_{k}) \nonumber \\
&\geq O(1/L)+\Delta_0\sum_{n\geq D}|z_{kn}|^2,
\end{align}
%%%%%%%%%%%%%%%%%%%%%%%%%
where $\Delta_0=E_D-E_{D-1}$.
This implies $\sum_{n\geq D}|z_{kn}|^2=O(1/L)$ for $k=0,\cdots,D-1$.
We also consider $\ket{\Phi_k'}=U^{-1}\ket{\Psi_{k}}=\sum_{n\geq0}(z^{-1})_{kn}\ket{\Psi_{n}}$ 
and its variational energy $\Delta E_k'=\sum_{n\geq0}|(z^{-1})_{kn}|^2(E_{n}-E_k)=O(1/L)$
for $k=0,\cdots,D-1$.
Then we obtain  $\sum_{n\geq D}|z^{\dagger}_{kn}|^2=\sum_{n\geq D}|z_{nk}|^2=O(1/L)$. %for $k=0,\cdots,D-1$.
%from which we have
%%%%%%%%%%%%%%%%%%%%%%%%%%%
%\begin{align}
%\left|\sum_{n\geq1}z_{kn}^*z_{k'n}\right|\leq \sum_{n\geq1}|z_{kn}^*z_{k'n}|=O(1/L).
%\end{align}
%%%%%%%%%%%%%%%%%%%%%%%%%
Therefore, the unitary matrix $z_{kk'}$ is almost block-diagonal and 
the submatrix $(Z)_{kk'}=z_{kk'}$ $(k,k'=0,\cdots,D-1)$ is almost a $D\times D$ unitary matrix
which satisfies
%%%%%%%%%%%%%%%%%%%%%%%%%%%
\begin{align}
\sum_{l=0}^{D-1}z_{kl}^*z_{k'l}=\delta_{kk'}+O(1/L).
\end{align}
%%%%%%%%%%%%%%%%%%%%%%%%%

%%%%%%%%%%%%%%%%%%%%%%%%%%%%%%%%%%%%%%%%%%%%%
{\it Proof of Claim~\ref{claim:Uq}.}
When the system has the translation symmetry, the ground states are simultaneous eigenstates of $T$
whose eigenvalues are shifted by application of $U$ (Eq.~\eqref{eq:UT}).
We can construct variational states $\ket{\Psi_1'}=U\ket{\Psi_{0}},\cdots,
\ket{\Psi_{q-1}'}=U^{q-1}\ket{\Psi_{0}}$ which have the variational energies $E_{0}+O(1/L)$.
Equation~\eqref{eq:UT} implies that $\ket{\Psi_{0}}$ and $\{\ket{\Psi_k'}\}$ have the $T$-eigenvalues, 
$\{1, e^{i2\pi\nu},e^{i4\pi\nu},\cdots,e^{i2\pi\nu(q-1)}\}$.
(We have assumed that $T\ket{\Psi_{0}}=\ket{\Psi_{0}}$ without generality.)
Therefore, $\ket{\Psi_{0}}$ and $\ket{\Psi_k'}$ have almost the same energies and 
are orthogonal each other.
Since the system is gapped with the degeneracy $D=q$, in the thermodynamic limit $L\to\infty$,
these variational states converge to each of the degenerate
ground states $\ket{\Psi_{k}}$ (up to phase factors) with the corresponding $T$-eigenvalues.
By the same reason, $U^q\ket{\Psi_{k}}$ converges to $\Psi_{k}$ itself up to a phase factor,
where they have the same momentum eigenvalue $e^{i2\pi\nu k}$.
The previous argument implies that the deviation of the matrix $\braket{\Psi_{k}|U^q|\Psi_{k'}}$ from 
the identity matrix is $O(1/L)$.
%The above discussions may be clear if one notices that $T$ and $U$ can be approximately 
%represented as Eq.\eqref{eq:TUmat}
%in the ground state sector with small deviations $O(1/L)$.

%%%%%%%%%%%%%%%%%%%%%%%%%%%%%%%%%%%%%%%%%%%%%
{\it Proof of Claim~\ref{claim:Uq_gapless}.}
Compared to gapped systems where the energy gap controls low energy properties,
it is difficult to analyze gapless systems.
Here, we provide an intuitively clear and simple discussion based on the gaplessness and the translation symmetry.
Although the twist operator itself acts within a fixed U(1) sector, the proof crucially relies on the response of the ground state energy to changes in the conserved charge. In particular, the vanishing inter-sector (charge) gap ensures smoothness of ground state expectation values across neighboring charge sectors, which ultimately constrains the intra-sector expectation value of $U^q$.
As stated in the condition of Claim~\ref{claim:Uq_gapless}, we assume that for a finite $L$, the ground state (more precisely, the ground state for the particle number $N$ and $N\pm1$) is
non-degenerate.

A gapless system is supposed to have a vanishing charge gap 
in the thermodynamic limit,
%%%%%%%%%%%%%%%%%%%%%%%%%%%
\begin{align}
\Delta_c^N\equiv E^{N+1}_0+E^{N-1}_0-2E_0^N \to 0,
\label{eq:gap_c}
\end{align}
%%%%%%%%%%%%%%%%%%%%%%%%%
where $E_0^N$ is the ground state energy in the $N$-particle sector.
In such a system, we naively expect that
%%%%%%%%%%%%%%%%%%%%%%%%%%%
\begin{align}
\braket{\Psi_0^{N+1}|{\mathcal O}|\Psi_0^{N+1}} +\braket{\Psi_0^{N-1}|{\mathcal O}|\Psi_0^{N-1}}
-2\braket{\Psi_0^{N}|{\mathcal O}|\Psi_0^{N}}\to 0
\label{eq:O}
\end{align}
%%%%%%%%%%%%%%%%%%%%%%%%%
for an hermitian operator ${\mathcal O}$ with a bounded norm.
Indeed, if we consider a perturbed Hamiltonian
%%%%%%%%%%%%%%%%%%%%%%%%%%%
\begin{align}
\tilde{H}=H+\lambda {\mathcal O},
\end{align}
%%%%%%%%%%%%%%%%%%%%%%%%%
change of the ground state energy is at most $\lambda|{\mathcal O}|$.
For a sufficiently small $\lambda$ in a finite (but large) system, 
there will be no level crossing between any energy levels and the system will stay in the gapless phase in the presence of 
the perturbation.
This implies
%%%%%%%%%%%%%%%%%%%%%%%%%%%
\begin{align}
\tilde{\Delta}_c^N\equiv \tilde{E}^{N+1}_0+\tilde{E}^{N-1}_0-2\tilde{E}_0^N \to 0
\end{align}
%%%%%%%%%%%%%%%%%%%%%%%%%
for the ground state energy $\tilde{E}_0^N$ of $\tilde{H}$.
%By differentiating the above equation with respect to $\lambda$ and setting $\lambda=0$,
%we obtain Eq.~\eqref{eq:O}.
Since the finite-size ground state is nondegenerate, we may differentiate the eigenvalue equation with respect to $\lambda$
and apply the Hellmann–Feynman theorem to obtain Eq.~\eqref{eq:O}.
%\textcolor{red}{
%Using the variational principle, we obtain a Hellmann–Feynman–type relation and the resulting equation~\eqref{eq:O},
%which is valid for a non-degenerate ground state for a finite $L$.
%Note that, throughout the proof, we first take the directional derivative at $\lambda=0$
%for each finite $L$, and only after that we take the thermodynamic limit $L\to \infty$.}
Now we consider two hermitian operators ${\mathcal O}_1=(U^q+U^{-q})$ and
${\mathcal O}_2=i(U^q-U^{-q})$ whose norms are $O(1)$.
Based on the above argument, by taking the sum $\braket{{\mathcal O}_{1}}-i\braket{{\mathcal O}_{2}}$ for $N$-, $N\pm1$-particle ground states,
we find that 
%%%%%%%%%%%%%%%%%%%%%%%%%%%
\begin{align}
2\braket{\Psi_0^{N}|U^q|\Psi_0^{N}} \simeq
\braket{\Psi_0^{N+1}|U^q|\Psi_0^{N+1}} +\braket{\Psi_0^{N-1}|U^q|\Psi_0^{N-1}}
\end{align}
%%%%%%%%%%%%%%%%%%%%%%%%%
for a large system.
When $\nu=N/L=p/q$ with $p,q$ coprime,
the fillings for the $(N\pm 1)$-particle states
$(N\pm 1)/L\equiv P_{\pm}/Q_{\pm}\neq p/q$ will 
have larger denominators $Q_{\pm}>q$.
Therefore, $\braket{\Psi_0^{N\pm1}|U^q|\Psi_0^{N\pm1}}=0$ by the translation symmetry,
which implies $\braket{\Psi_0^{N}|U^q|\Psi_0^{N}}=0$.
%If the ground state is degenerate for finite $L$, the argument can be reformulated in terms of the ground-state projector.
Besides,
the proof naturally suggests an algebraic decay of $|\braket{U^q}|$ with the system size $L$ for gapless states,
since the charge gap often displays an algebraic size dependence.
%according to our proof, the inter-sector charge gap often displays an algebraic size dependence, which in turn suggests a power-law behavior of $|\braket{U^q}|$ with respect to $L$.

%%%%%%%%%%%%%%%%%%%%%%%%%%%%%%%%%%%%%%%%%%%%%%%%%%%%%%%%%%%%
\section{higher dimensional systems}
\label{sec:highD}
%%%%%%%%%%%%%%%%%%%%%%%%%%%%%%%%%%%%%%%%%%%%%%%%%%%%%%%%%%%%

It is known that a straightforward generalization of the twist operator $U$ does not work 
in higher dimensions~\cite{LSM1961,WatanabeOshikawa2018,Oshikawa2000_1,Kang2019,Wheeler2019,Ono2019}.
Nevertheless,
our arguments on one-dimensional systems can be generalized to higher dimensions.
%In doing so, 
In the previous sections,
the explicit definitions of $T$ and $U$ (Eqs.~\eqref{eq:T} and ~\eqref{eq:U}) themselves were not essential,
but the two properties, (i) their commutation relation and (ii) creation of low energy states by $U$,
played central roles.
This suggests that we can repeat the same argument for higher dimensions, 
if there are suitable operators with these properties corresponding 
to $T$ and $U$.

To be concrete, we consider a two dimensional square lattice system
described by Eq.~\eqref{eq:H} with the size 
$L_x\times L_y$ $(L_x\simeq L_y\simeq L)$ under the
periodic boundary conditions in both directions, where
a site position is denoted as $j=(x_j,y_j)$.
The filling is $\nu=N/(L_xL_y)=p/q$ with $p,q$ coprime.
We suppose that the system has time-reversal symmetry in addition to the U(1) and translation symmetries 
(more precisely, the Hall conductivity is assumed to be zero, which is valid for non-Chern bands).
%Other lattices and three dimensional systems can be treated in a similar manner~\cite{}.
To control the system, 
we introduce a uniform background U(1) flux so that each plaquette acquires 
a flux $\phi=2\pi/(L_xL_y)$~\cite{Tada2021,Tada2025}.
Specifically, we use the gauge configuration $A_{ij}=A_{ij}^x+A_{ij}^y$ with
$A_{ij}^x=-\phi L_x(y_i+y_j)\delta_x/2, A_{ij}^y=\phi(x_i+x_j)(y_i-y_j)/2$,
where $\delta_{x}=\pm1$ for the hopping between $x=0$ and $L_x-1$ and $\delta_x=0$ otherwise.
One can introduce such gauge fields for general lattices~\cite{Tada2025}.
It was shown that there is no exact magnetic translation symmetry in this system and the translation symmetry 
is approximate~\cite{Tada2021}.
Then, an approximate variant of 
the magnetic translation operators $\mathcal{T}_{x,y}$ has similar roles as those of $T, U$.
They are given by
%%%%%%%%%%%%%%%%%%%%%%%%%%%
\begin{align}
\mathcal{T}_x&=T_xe^{i\phi\sum_jY_jn_j}, \quad 
\mathcal{T}_y=T_ye^{i\phi\sum_jX_jn_j},
\end{align}
%%%%%%%%%%%%%%%%%%%%%%%%%
where $T_{x,y}$ are the standard translation operators and $Y_j=L_x(L_y-y_j)\delta_{x_j,L_x-1}, X_j=x_j$.
Note that $\phi\sum X_jn_j=(2\pi/L^2)\sum x_jn_j$ has a close connection to polarization for two dimensional systems in the present gauge,
and indeed $\mathcal{T}_{x,y}$ can be regarded as polarization operators~\cite{Song2021}. 
Our construction should be viewed as a symmetry-based many-body index rather than a straightforward generalization of the microscopic polarization operator.

Let us denote $\mathcal{T}_x= \mathcal{T}$ and $\mathcal{T}_y^{}=\mathcal{U}$ 
so that analogy to one dimensional systems is clear. 
The two approximate symmetry operators satisfy the commutation relation
%%%%%%%%%%%%%%%%%%%%%%%%%%%
\begin{align}
\mathcal{U}_{}\mathcal{T}_{}\mathcal{U}_{}^{-1}=e^{i\phi N}\mathcal{T}_{},
\label{eq:UT2d}
\end{align}
%%%%%%%%%%%%%%%%%%%%%%%%%
where $\phi N=2\pi\nu$ in two dimensions. 
This commutation relation is gauge independent and therefore the following discussions essentially do not depend on 
gauge choices~\cite{Tada2025,Tada2023}. 
For three dimensional systems with filling $\nu=N/(L_xL_yL_z)=p/q$, 
we take the system size $L_z$
to be coprime with $q$ so that $\phi N=2\pi\nu L_z\neq 0$ (mod $2\pi$)~\cite{Oshikawa2000_1,Oshikawa2000_2},
where we suppose that the thermodynamic limit is well described by the chosen sequence $\{L_z\}$.
Generally, for two dimensions with the total number of the unit cells $L_xL_y$, 
we introduce the total flux $\Phi=\sum_p\phi_p=2\pi$ in the entire system, where the flux for each plaquette $\phi_p$ 
is $O(L^{-2})$. For three dimensions, the total flux on a two-dimensional cross section is $\Phi=2\pi$.
One can use a gauge field which is consistent with the underlying lattice~\cite{Tada2021,Tada2025}.
The system under the flux is approximately one unit cell translationally symmetric up to gauge transformation.
It is straightforward to construct the corresponding approximate magnetic translation operators by requiring
that changes of the gauge field under the standard translation are compensated by a gauge transformation~\cite{Tada2021,Tada2025}.
Then, we obtain the commutation relation Eq.~\eqref{eq:UT2d}  for general lattices.
Once the commutation relations have been obtained, we can repeat the same arguments to obtain the claims as follows.

By following the previous study~\cite{Tada2021} and the preceding sections, 
we can show that the variational state energy compared to the ground state energy is
%%%%%%%%%%%%%%%%%%%%%%%%%%%
\begin{align}
\Delta E_{k}=\braket{\Psi_{k}|\mathcal{U}_{}^{-1}H\mathcal{U}_{}-H|\Psi_{k}}=O(1/L),
\label{eq:dE2d}
\end{align}
%%%%%%%%%%%%%%%%%%%%%%%%%
where the Hamiltonian $H$ contains the uniform flux $\phi$.
%(we have neglected the energy difference within the quasi-degenerate ground states for simplicity).
%We have $\Delta E_k=O(1/L)$ in three dimensions.
This corresponds to Claim~\ref{claim:dE}.
With use of Eqs.~\eqref{eq:UT2d} and \eqref{eq:dE2d},
we can repeat the same argument as in the previous sections.
Then, we can derive Claim~\ref{claim:detZ} for the Hamiltonian 
with the flux $\phi$
under the same conditions.
%where $O(1/L)$ is replaced by $O(1/L^{4-d})$ with the dimensionality $d=2,3$.
Claims~\ref{claim:Uq} and ~\ref{claim:Uq_gapless} follow from the approximate $\mathcal{T}$-symmetry and 
the almost orthogonality of the variational states,
$|\braket{\Psi_{k}|\mathcal{U}^q|\Psi_{k'}}|=\delta_{kk'}+O(1/L)$~\cite{Tada2021}.
Therefore, our statements on the expectation values of $\mathcal{U}^q$ hold under the small flux in higher dimensions.
Finally, we can extrapolate these results to the original flux free Hamiltonian, since the flux 
$\phi=O(1/L^2)$ is negligibly small and the time-reversal symmetry guarantees that the energy spectra are 
perturbatively robust to the flux.
Indeed, it was shown that the spectra of the Hamiltonians with and without 
the flux are almost unchanged~\cite{Tada2021,Tada2025}.
Therefore, when the flux free Hamiltonian has gapped degenerate ground states,
the ground state expectation value of $\mathcal{U}^q$ under the flux $\phi$ is 
$|\braket{\Psi_0|\mathcal{U}^q|\Psi_0}|\to1$ 
in the thermodynamic limit.
Similarly, $|\braket{\Psi_0|\mathcal{U}^q|\Psi_0}|\to0$ under the flux, when the flux free Hamiltonian is gapless.

From the above argument, we conclude that
all of the Claims 1$\sim$4 holds even in higher dimensions with the replacement $U\to{\mathcal U}$ under the time-reversal symmetry. 
We can explicitly confirm these statements in concrete models 
as discussed in Appendix~\ref{app:2d}.

%%%%%%%%%%%%%%%%%%%%%%%%%%%%%%%%%%%%%%%%%%%%%%%%%%%%%%%%%%%%
\section{Summary and Discussion}
%%%%%%%%%%%%%%%%%%%%%%%%%%%%%%%%%%%%%%%%%%%%%%%%%%%%%%%%%%%%
We have developed the scheme to distinguish between gapless and gapped states based on the twist operator 
under the U(1) and translation symmetries.
We provided a general proof for the criteria that $|\braket{\Psi_0|U^q|\Psi_0}|\to0$ in a charge gapless system 
and $|\braket{\Psi_0|U^q|\Psi_0}|\to1$
in a gapped system with the filling $\nu=p/q$ in one dimension.
For higher dimensions, the approximate translation operator under a small flux is introduced and shown to be quantized in a similar manner. 
Our method will be practically useful in numerical calculations of finite size systems 
and also can provide an insight into
the characterization of gapped and gapless states based on symmetries.

The quantity $|\braket{\Psi_0|U^q|\Psi_0}|$ indeed works as a quantized order parameter or an index as proposed in the previous studies.
However, we require an additional assumption in Claim~\ref{claim:Uq} on the ground state degeneracy for a gapped system and thus we cannot exclude
a possibility that there is a gapped state with some degeneracy which has $|\braket{\Psi_0|U^q|\Psi_0}|<1$.
Similarly, the proof of Claim~\ref{claim:Uq_gapless} applies only to non-degenerate gapless ground states, although
the argument could be reformulated in terms of the ground state projector for a degenerate case.
For the higher dimensional generalization, our method is applicable only to two and three dimensions but not to four or higher dimensions.
Besides, the time-reversal symmetry is required and a system with anomalous Hall conductivity cannot be discussed.
These problems are left for future studies.

%%%%%%%%%%%%%%%%%%%%%%%%%%%%%%%%%%%%%%%%%%%%%%%%%%%%%%%%%%%%
\section*{Acknowledgements}
%%%%%%%%%%%%%%%%%%%%%%%%%%%%%%%%%%%%%%%%%%%%%%%%%%%%%%%%%%%%
We thank Shunsuke C. Furuya, Masaaki Nakamura, Masaki Oshikawa, Hal Tasaki, Haruki Watanabe, and Yuan Yao for fruitful discussions.
This work is supported by JSPS KAKENHI Grant No.
22K03513 and No. 26K06946.

%\bibliography{ref}

%%%%%%%%%%%%%%%%%%%%%%%%%%%%%%%%%%%%%%%%%%%%%%%%%%%%%%%%%

\appendix
%%%%%%%%%%%%%%%%%%%%%%%%%%%%%%%%%%%%%%%%%%%%%%%%%%%%%%%%%%%

%%%%%%%%%%%%%%%%%%%%%%%%%%%%%%%%%%%%%%
\section{Numerical demonstration}
%%%%%%%%%%%%%%%%%%%%%%%%%%%%%%%%%%%%%
%%%%%%%%%%%%%%%%%%%%%%%%%%%%%%%%%%%%%%%%%%%%%%%%%%%%%%%%%%%%
\subsection{One-dimensional system}
\label{app:1d}
%%%%%%%%%%%%%%%%%%%%%%%%%%%%%%%%%%%%%%%%%%%%%%%%%%%%%%%%%%%%

We consider interacting fermions in one-dimension,
%%%%%%%%%%%%%%%%%%%%%%%%%%%
\begin{align}
H=\sum_{i=0}^{L-1}tc^{\dagger}_ic_{i+1}+{\rm (h.c.)} + \sum_{i,k}V_{k} n_i n_{i+k},
\end{align}
%%%%%%%%%%%%%%%%%%%%%%%%%
where $L$ is assumed to be $L\in3\mathbb{Z}$.
We suppose that
the filling is $\nu=N/L=1/3$ and the interaction is $V_1=V_2=V$ and $V_k=0$ otherwise.
The energy unit is $t=1$.
We introduce a twisted boundary condition with a sufficiently small twist 
%so that the gapless ground state is non-degenerate and the condition of Claim~\ref{claim:Uq_gapless} is satisfied. 
so that possible exact degeneracy in the gapless ground state is lifted and the condition of Claim~\ref{claim:Uq_gapless} is satisfied.
(Claim~\ref{claim:Uq_gapless} holds also for systems with the twisted boundary condition.)

In the non-interacting limit $V=0$, the ground state is simply given by the Fermi sea state
%%%%%%%%%%%%%%%%%%%%%%%%%%%
\begin{align}
\ket{\Psi_0}=\prod_{\ell}\hat{c}_{\ell}^{\dagger}\ket{0},
\end{align}
%%%%%%%%%%%%%%%%%%%%%%%%%
where $\hat{c}_{\ell}$ is the Fourier transformation of $c_j$ and the lowest $N$ single-particle levels are occupied.
It is clear that $\braket{\Psi_0|U^k|\Psi_0}=0$ for $k=1,2,3$ due to the translation symmetry,
which is trivially consistent with Claim~4. %\ref{claim:Uq_gapless}.
In the strong coupling limit $V/t\to\infty$,
the ground states are three fold degenerate ($D=3$),
%%%%%%%%%%%%%%%%%%%%%%%%%%%
\begin{align}
\ket{\Psi_n'}=\prod_jc_{n+3j}^{\dagger}\ket{0}
\end{align}
%%%%%%%%%%%%%%%%%%%%%%%%%
with $n=0,1,2$. There is an excitation gap $\Delta_0\sim V$ above the degenerate ground states.
Under the translation operation, $T\ket{\Psi_n'}=\ket{\Psi_{n+1}'}$ (mod 3).
We can construct translationally symmetric ground states
%%%%%%%%%%%%%%%%%%%%%%%%%%%
\begin{align}
\ket{\Psi_n}=\frac{1}{\sqrt{3}}\sum_{m=0}^2e^{-i2\pi nm/3}\ket{\Psi_m'}
\end{align}
%%%%%%%%%%%%%%%%%%%%%%%%%
which satisfy $T\ket{\Psi_n}=e^{i2\pi\nu n}\ket{\Psi_n}$.

It is straightforward to calculate overlap matrix in the gapped phase with the above wavefunctions,
%%%%%%%%%%%%%%%%%%%%%%%%%%%
\begin{align}
\braket{\Psi'_m|U|\Psi'_n}&=
\begin{bmatrix}
   e^{i\pi} & 0 & 0\\
   0 & e^{-i\pi/3} &0 \\
   0 &  0 & e^{i\pi/3}
\end{bmatrix}_{mn},\\
\braket{\Psi_m|U|\Psi_n}&=
\begin{bmatrix}
   0 & -1 & 0\\
   0 & 0 &-1 \\
   -1 &  0 & 0
\end{bmatrix}_{mn},
\end{align}
%%%%%%%%%%%%%%%%%%%%%%%%%
for which $|\det(Z)|=1$ consistently with Claim~2. % \ref{claim:detZ}.
Since the ground state degeneracy is $D=q=3$ in the gapped phase, the condition of Claim~3. %\ref{claim:Uq}
is satisfied.
Indeed, we have $|\braket{\Psi'_m|U^3|\Psi'_n}|=|\braket{\Psi_m|U^3|\Psi_n}|= \delta_{mn}$.

\begin{figure}[htb]
\includegraphics[width=0.95\columnwidth]{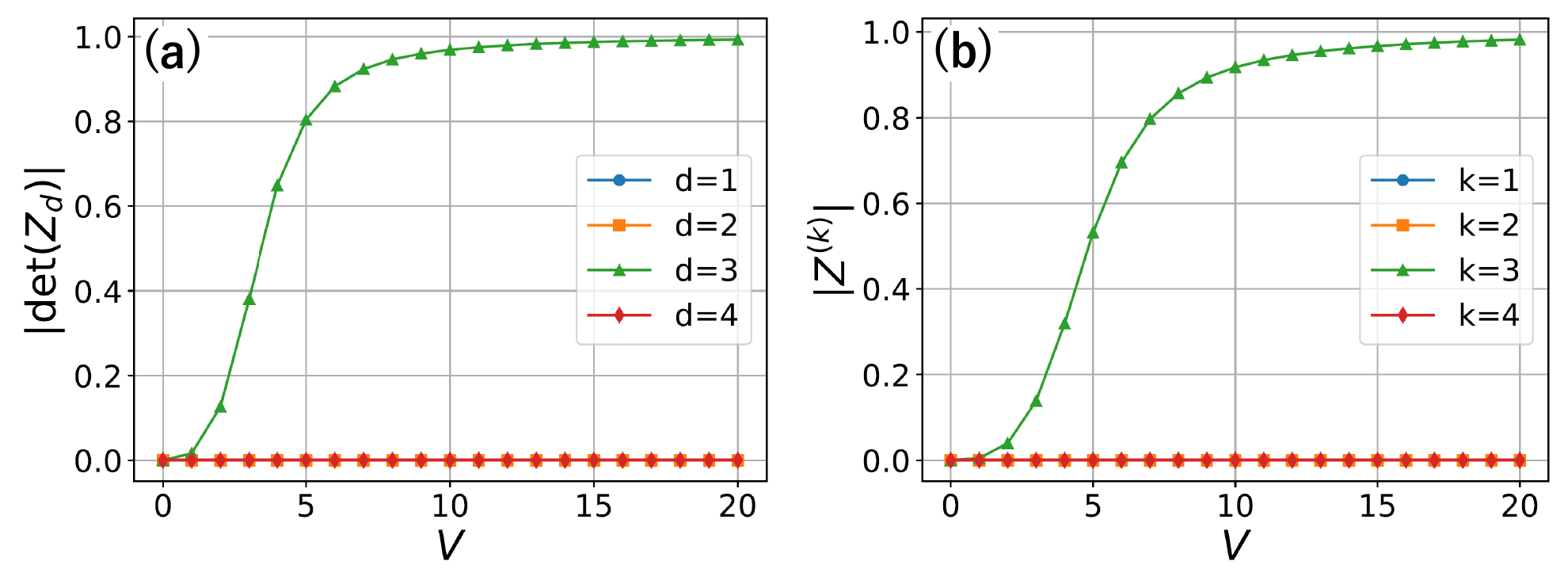}
\caption{(a) $|\det (Z_d)|$ and (b) $|Z^{(k)}|$ at the filling $\nu=1/3$ for the small system size $N=5, L=15$.
The curves for $d=1,2,4$ and $k=1,2,4$ almost coincide in each case.
}
\label{fig:1dZ}
\end{figure}

For general parameters,
we perform numerical exact diagonalization of a small size system, $N=5, L=15$.
%We show energy spacing $\Delta E_{k0}=E_k-E_0$ between the ground state energy and the $k$-th energy
%in Fig.~\ref{fig:1dE}.
%$\Delta E_{10}$ is vanishingly small for all $V$ due to the inversion symmetry of the system.
%$\Delta E_{20}$ approaches zero and quasi three-fold degeneracy is seen in the  strong interaction region.
We define $d\times d$ matrices by using the lowest $d$ eigenstates,
%%%%%%%%%%%%%%%%%%%%%%%%%%%
\begin{align}
(Z_d)_{mn}=\braket{\Psi_m|U|\Psi_n},
\end{align}
%%%%%%%%%%%%%%%%%%%%%%%%%
where $m,n=0,1,\cdots,d-1$.
As seen in Fig.~\ref{fig:1dZ} (a), $|\det(Z_d)|=0$ for all $d$ for small interactions where the system is gapless.
They remain zero except for $d=3$ even in the strong interaction regime, 
while $|\det(Z_3)|\to1$ for $d=3$.
These behaviors are consistent with Claim~2. %\ref{claim:detZ}.
We also consider 
%%%%%%%%%%%%%%%%%%%%%%%%%%%
\begin{align}
Z^{(k)}=\braket{\Psi_0|U^k|\Psi_0}.
\end{align}
%%%%%%%%%%%%%%%%%%%%%%%%%
As shown in Fig.~\ref{fig:1dZ} (b), $|Z^{(k)}|$ shows similar behaviors to those of $|\det(Z_d)|$.
$|Z^{(k)}|\simeq0$ for all $k$ in the gapless region and only $|Z^{(3)}|$ approaches $|Z^{(3)}|\to1$
in the gapped regime, which is consistent with Claims 3 and 4. %~\ref{claim:Uq} and \ref{claim:Uq_gapless}.

%%%%%%%%%%%%%%%%%%%%%%%%%%%%%%%%%%%
%\begin{figure}[htb]
%\includegraphics[width=5.0cm]{fig/1d_detZd_L15.pdf}
%\caption{$|\det (Z_d)|$ at the filling $\nu=1/3$ for the small system size $N=5, L=15$.
%The curves for $d=1,2,4$ almost coincide.
%}
%\label{fig:1dDetZ}
%\end{figure}
%%%%%%%%%%%%%%%%%%%%%%%%%%%%%%%%%%%%
%%%%%%%%%%%%%%%%%%%%%%%%%%%%%%%%%%%
%\begin{figure}[htb]
%\includegraphics[width=5.0cm]{fig/1d_Zk_L15.pdf}
%\caption{$|Z^{(k)}|$ at the filling $\nu=1/3$ for the small system size $N=5, L=15$.
%The curves for $d=1,2,4$ almost coincide.
%}
%\label{fig:1dZk}
%\end{figure}
%%%%%%%%%%%%%%%%%%%%%%%%%%%%%%%%%%%%

\begin{figure}[htb]
\includegraphics[width=0.95\columnwidth]{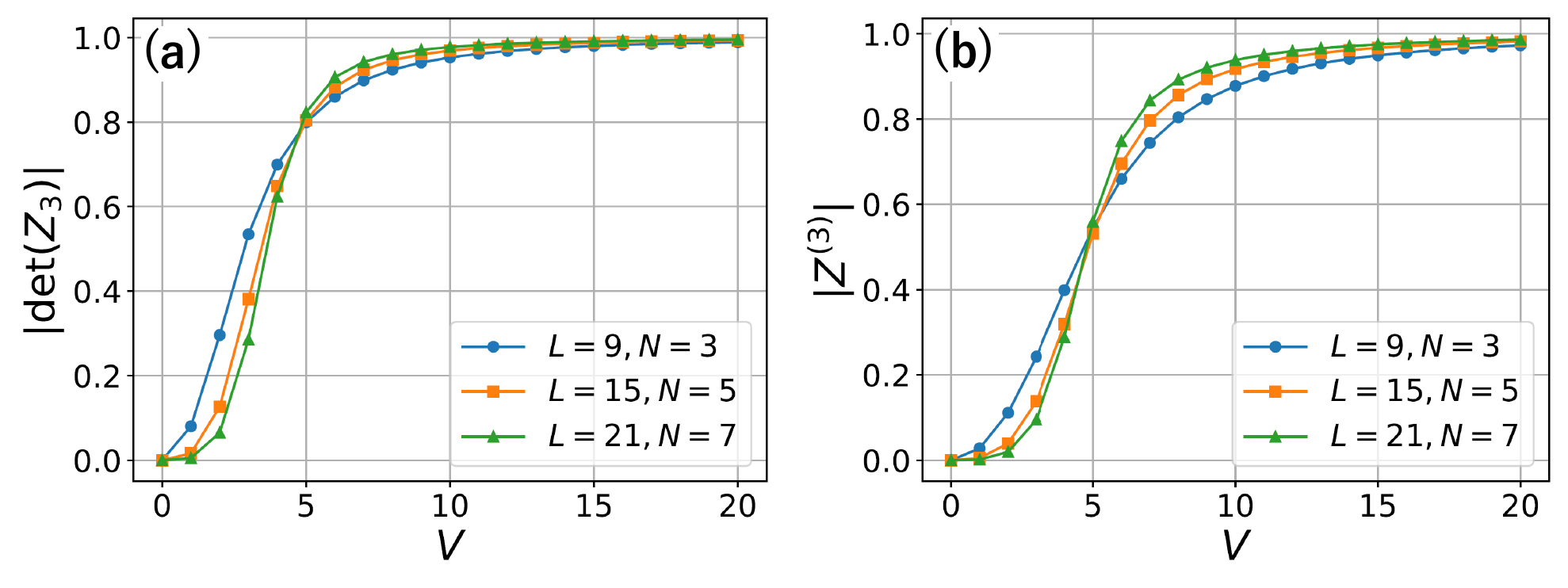}
\caption{(a) $|\det (Z_d)|$ and (b) $|Z^{(k)}|$ at the filling $\nu=1/3$ for the system sizes $L=9, 15, 21$.
}
\label{fig:1dZ3}
\end{figure}

There are finite size effects in $|\det(Z_3)|$ and $|Z^{(3)}|$ for small system sizes.
We show $|\det(Z_3)|$ and $|Z^{(3)}|$ for $L=9, 15, 21$ in Figs.~\ref{fig:1dZ3}, respectively.
As the system size increases,
they decrease for the gapless region $V\leq V_c$ and approaches unity in the gapped region $V>V_c$,
where the transition point is located around $V_c\simeq 5t$.
In the thermodynamic limit, they will be discontinuous at $V=V_c$ and take only the quantized values $\{0,1\}$,
although it is difficult to perform accurate finite size scaling with using only the small system sizes $L=9, 15, 21$.
Systematic numerical investigation will be an interesting future study.

%%%%%%%%%%%%%%%%%%%%%%%%%%%%%%%%%%%
%%%%%%%%%%%%%%%%%%%%%%%%%%%%%%%%%%%%

%%%%%%%%%%%%%%%%%%%%%%%%%%%%%%%%%%%
%%%%%%%%%%%%%%%%%%%%%%%%%%%%%%%%%%%%

%%%%%%%%%%%%%%%%%%%%%%%%%%%%%%%%%%%
%\begin{figure}[htb]
%\includegraphics[width=5.0cm]{fig/detZ3.pdf}
%\caption{$|\det (Z_3)|$ at the filling $\nu=1/3$ for $L=9, 15, 21$.
%}
%\label{fig:detZ3}
%\end{figure}
%%%%%%%%%%%%%%%%%%%%%%%%%%%%%%%%%%%%
%%%%%%%%%%%%%%%%%%%%%%%%%%%%%%%%%%%
%\begin{figure}[htb]
%\includegraphics[width=5.0cm]{fig/Z3.pdf}
%\caption{$|Z^{(3)}|$ at the filling $\nu=1/3$ for $L=9, 15, 21$.
%}
%\label{fig:Z3}
%\end{figure}
%%%%%%%%%%%%%%%%%%%%%%%%%%%%%%%%%%%%

%%%%%%%%%%%%%%%%%%%%%%%%%%%%%%%%%%%%%%%%%%%%%%%%%%%%%
\subsection{Two-dimensional system}
\label{app:2d}
%%%%%%%%%%%%%%%%%%%%%%%%%%%%%%%%%%%%%%%%%%%%%%%%%%%%%%%

%%%%%%%%%%%%%%%%%%%%%%%%%%%%%%%%%%%
%\begin{figure}[htb]
%\includegraphics[width=5.0cm]{fig/2dv.pdf}
%\caption{Numerical results of $\Delta_0$, $U_1$, and $U_2$ for two system sizes $L=10,20$.
%The energy unit is $t=1$.
%}
%\label{fig:2dv}
%\end{figure}
%%%%%%%%%%%%%%%%%%%%%%%%%%%%%%%%%%%%
%%%%%%%%%%%%%%%%%%%%%%%%%%%%%%%%%%%
%\begin{figure}[htb]
%\includegraphics[width=5.0cm]{fig/2dL.pdf}
%\caption{Numerical results of $\Delta_0$, $U_1$, and $U_2$ at $v=2t$.
%The energy unit is $t=1$.
%}
%\label{fig:2dL}
%\end{figure}
%%%%%%%%%%%%%%%%%%%%%%%%%%%%%%%%%%%%

We consider free fermions with filling $\nu=N/(L_xL_y)=1/2$ per site
on the two dimensional square lattice with $L_x=L_y=L\in 2\mathbb{Z}$.
The Hamiltonian is 
%%%%%%%%%%%%%%%%%%%%%%%%%%%
\begin{align}
H=\sum_{ij}te^{iA_{ij}}c^{\dagger}_ic_j + \sum_iv_ic^{\dagger}_ic_i ,
\end{align}
%%%%%%%%%%%%%%%%%%%%%%%%%
where $t=1$ is the nearest neighbor hopping and $v_i=v(-1)^{|i|}$ is the staggered on-site potential.
$A_{ij}$ is the vector potential which gives a uniform flux $\sum_{ij\in{\rm plaquette}}A_{ij}=\phi=2\pi/L^2$.
The ground state is a Fermi sea metal for $v=0$, while it is a charge-density-wave band insulator for $v\neq0$.
The system has the approximate magnetic translation symmetry 
$\mathcal{T}^2\equiv\mathcal{T}_x^2, \mathcal{U}^2\equiv\mathcal{T}_y^2$ for general $v\neq0$,
where $\mathcal{T}_{x,y}$ correspond to one site translation.
(The unit cell contains two sites when $v\neq0$ and the filling per unit cell is $\nu=1$.
One can discuss the system also based on this notation.)
For the metallic state, it is difficult to analytically evaluate $\braket{\mathcal{U}^2}$ in the presence of $\phi$.
On the other hand, in the deep insulating region $v/t\to\infty$,
charge distribution is uniquely fixed as $n_j=(-1)^{|j|}/2$ and the ground state expectation 
$\braket{\Psi_0|\mathcal{U}^2|\Psi_0}$ is easily calculated.
Since $\mathcal{U}^2=(T_ye^{i\phi\sum_jx_jn_j})^2=T_y^2e^{i2\phi\sum_jx_jn_j}$ and $T_y^2\ket{\Psi_0}=\ket{\Psi_0}$
in the insulating limit,
we obtain
%%%%%%%%%%%%%%%%%%%%%%%%%%%
\begin{align}
|\braket{\Psi_0|\mathcal{U}^2|\Psi_0}|=1.
\end{align}
%%%%%%%%%%%%%%%%%%%%%%%%%

\begin{figure}[htb]
\includegraphics[width=0.95\columnwidth]{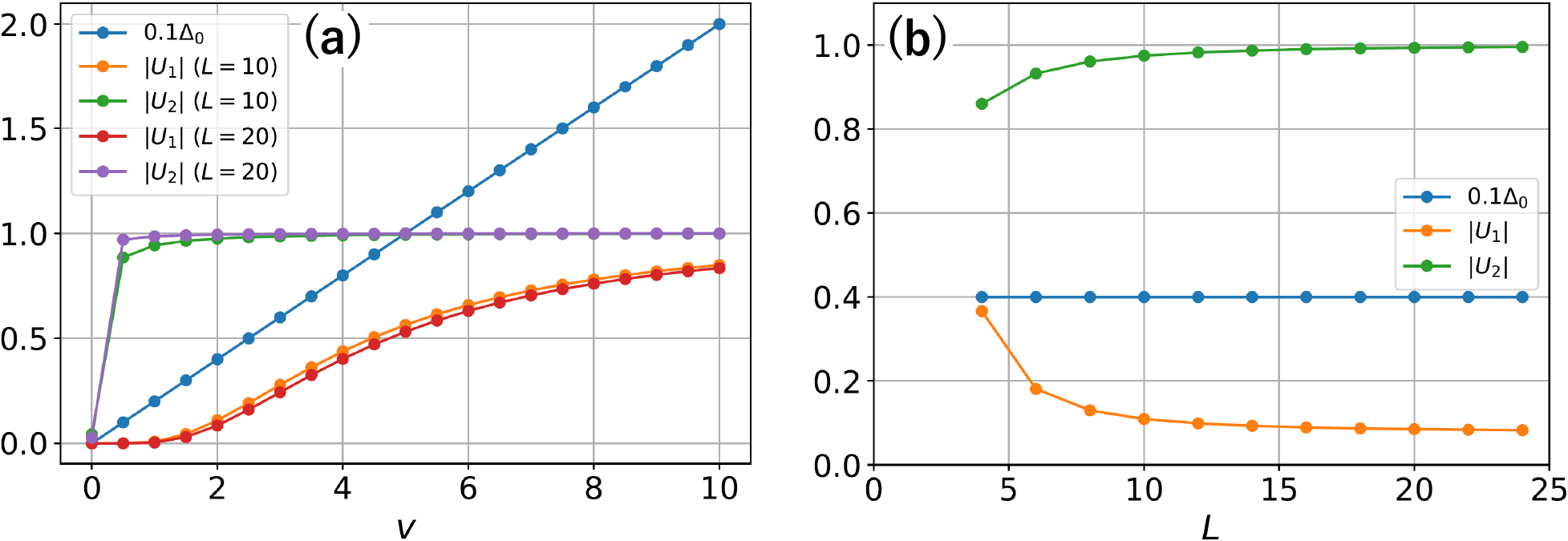}
\caption{Numerical results of $\Delta_0$, $U_1$, and $U_2$ (a) for two system sizes $L=10,20$ and (b) for the potential $v=2t$.
}
\label{fig:2dvL}
\end{figure}

\begin{figure}[htb]
\includegraphics[width=5.0cm]{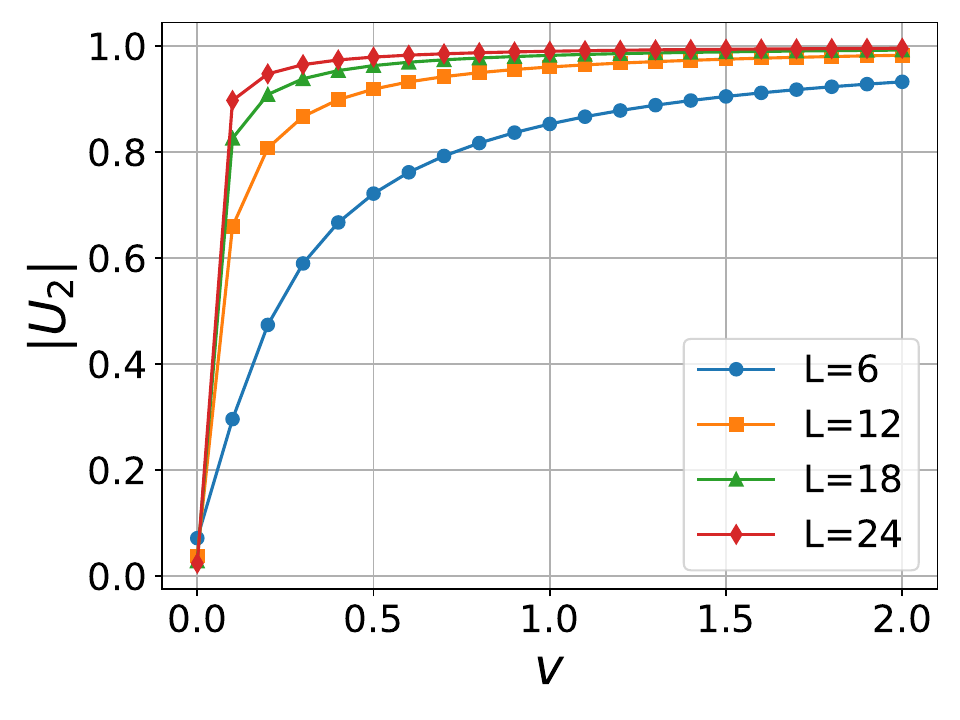}
\caption{Numerical results of $U_2$ for a small $v$ region.
}
\label{fig:2dvsmall}
\end{figure}

We can numerically evaluate $\braket{\mathcal{U}^2}$ under the flux $\phi$ for general parameters.
The excitation gap is defined as $\Delta_0=E_1-E_0=\varepsilon_{N}-\varepsilon_{N-1}$,
where $\varepsilon_n$ $(n=0,1,\cdots)$ are the single-particle energies.
We compute two quantities
%%%%%%%%%%%%%%%%%%%%%%%%%%%
\begin{align}
U_1= \braket{\Psi_0|U^2|\Psi_0},\quad U_2= \braket{\Psi_0|\mathcal{U}^2|\Psi_0},
%U_1\equiv \braket{\Psi_0|e^{i4\pi/L\sum_jx_jn_j}|\Psi_0},\\
%U_2\equiv \braket{\Psi_0|T_ye^{i2\phi\sum_jx_jn_j}|\Psi_0}.
\label{eq:U1U2}
\end{align}
%%%%%%%%%%%%%%%%%%%%%%%%%
where $U=e^{i2\pi/L\sum_jx_jn_j}$ is the polarization operator straightforwardly extended to two dimensions.
$U_1$ is the previously proposed order parameter for one-dimensional systems~\cite{RestaSorella1999,
AligiaOrtiz1999},
and $U_2$ is the index defined by the approximate magnetic translation operator.
($\braket{U}$ and $\braket{U^2}$ behave in a qualitatively similar manner. 
Here, we consider the latter one for a comparison with $\braket{\mathcal{U}^2}$.)
Note that the ground state becomes an eigenstate for both operators in the limit $v/t\to \infty$,
and thus $|U_1|, |U_2|\to1$ in this limit. 
As shown in Fig.~\ref{fig:2dvL} (a),
both of them are vanishing in the metallic state with $v=0$.
$|U_1|$ remains nearly zero up to $v=t\sim2t$ and approaches $|U_1|=1$ slowly, 
but it does not reach $|U_1|=1$ even for a large $v=10t$.
Besides, it decreases for larger system sizes and approaches some non-universal 
values depending on $v$ in the thermodynamic limit as seen in Fig.~\ref{fig:2dvL} (b).
In three dimensions, it will vanish for large system sizes.
Therefore, $U_1$ is not a quantized order parameter for gapless and gapped states in higher dimensions.
On the other hand, $U_2$ quickly converges to $|U_2|\to1$ in the insulating regime $v>0$,
and it increases toward $|U_2|\to1$ for larger system sizes.
These behaviors of $U_2$ are consistent with the discussions in the main text
(Claims 3 and 4). %~\ref{claim:Uq} and \ref{claim:Uq_gapless}).
We stress that $|U_2|\to1$ for large $L$ as long as $v\neq0$ (namely $\Delta_0\neq0$)
as seen in Fig.~\ref{fig:2dvsmall}.
This is in sharp contrast to $|U_1|$ which is numerically very small for $v\lesssim t$ (Fig.~\ref{fig:2dvL} (a)).

%%%%%%%%%%%%%%%%%%%%%%%%%%%%%%%%%%%
%%%%%%%%%%%%%%%%%%%%%%%%%%%%%%%%%%%%
%%%%%%%%%%%%%%%%%%%%%%%%%%%%%%%%%%%
%%%%%%%%%%%%%%%%%%%%%%%%%%%%%%%%%%%%

\bibliography{ref}

%%%%%%%%%%%%%%%%%%%%%%%%%%%%%%%%%%%%%%%%%%%%%%%%%%%%%%%%%%%%%%%%%%%%%%%%
%%%%%%%%%%%%%%%%%%%%%%%%%%%%%%%%%%%%%%%%%%%%%%%%%%%%%%%%%%%%%%%%%%%%%%%%

\end{document}